\documentclass[twoside,12pt]{article}  
\usepackage{indentfirst} 
\usepackage{bm} 
\usepackage{graphicx} 
\usepackage{amsmath} 

\begin{document} 


\thispagestyle{empty} \vspace*{0.8cm}\hbox
to\textwidth{\vbox{\hfil\huge\sf Chinese Physics B\hfill}\hfill}
\par\noindent\rule[3mm]{\textwidth}{0.2pt}\hspace*{-\textwidth}\noindent
\rule[2.5mm]{\textwidth}{0.2pt}


\begin{center}
\LARGE\bf Comment on `Mathematical structure of the three-dimensional (3D) Ising model'$^{*}$    
\end{center}

\begin{center}  
Jacques H.H.\ Perk$^{\dagger}$
\end{center}

\begin{center}  
\begin{small} \sl
Department of Physics, Oklahoma State University,
Stillwater, OK 74078-3072, USA\\   
\end{small}
\end{center}

\begin{center}  
\small (Received 20 June 2013)
\end{center}

\vspace*{2mm}

\begin{center}  
\begin{minipage}{15.5cm}
\parindent 20pt\small
The review paper by Zhang Zhi-Dong contains many errors and is based on
several earlier works that are equally wrong.

\end{minipage}
\end{center}

\begin{center}  
\begin{minipage}{15.5cm}
\begin{minipage}[t]{2.3cm}{\bf Keywords:}\end{minipage}
\begin{minipage}[t]{13.1cm}
Ising model, Lie algebra, series analysis, thermodynamic limit
\end{minipage}\par\vglue8pt
{\bf PACC: }
05.50+q, 75.10.Hk, 05.70.Fh
\end{minipage}
\end{center}

\footnotetext[1]{Project supported in part by NSF grant PHY 07-58139. }
\footnotetext[2]{Corresponding author. E-mail:  perk@okstate.edu}


\newcommand{\bG}{\boldsymbol{\mathit{\Gamma}}}
\newcommand{\Vb}{\boldsymbol{\mathit{V}}}
\newcommand{\Tb}{\boldsymbol{\mathit{T}}}

\section{Introduction}  

In 2007 Zhi-Dong Zhang published the very long paper \cite{Z}, claiming to
present complete exact solutions for the free energy and spontaneous
magnetization per site of the Ising model on a three-dimensional
orthorhombic lattice. He also claimed results for the pair correlation function.
That his claims are false is clearly stated in \cite{Wu1,Perk1}. Zhang's formulae
(74)  and (102) with his choice of the weights $w_x=1$, $w_y=w_z=0$,
see pages 5339 and  5370, do not reproduce the well-established high- and
low-temperature series results, as required by rigorous theorems to the
contrary. Also his paper \cite{Z} opens with an incorrect application of the
Jordan--Wigner transformation \cite{Perk1}.

To this criticism Zhang brings up  ``the possibility of the occurrence of a
phase transition at infinite temperature according to the Yang--Lee theorems''
on page 3097 of \cite{Z2}, see also pages 5369--5371 of \cite{Z}. This is a
clear error, as Zhang's picture means that the partition function zeros in the
large system limit are to pinch the real $\beta$ axis for $\beta=0$ and
$\beta_{\mathrm{c}}$, but not in between. Thus he claims a singularity
at $\beta=0$, even for zero magnetic field, such that one cannot apply
the series test. This violates rigorous theorems. The absurdity of this
argument is immediately clear as Zhang's formula (49) in \cite{Z} can be
expanded in $\beta$ with a finite radius of convergence, so that he contradicts
himself. Pinching of Yang--Lee zeros at $\beta=0$ would require zero
radius of convergence.

In spite of the fact that Zhang's magnum opus is clearly in error, he published
many more papers, mostly with Norman H.~March, adding more errors.
The last paper \cite{ZM}, to which this comment applies, contains all relevant
references, so that the referees of this work are without excuse for failing
to reject it.

In the next few sections we shall discuss several of the errors in Zhang's
work in more detail.


\section{Zhang's results violate established series results,
even in first nontrivial orders}

To explicitly see what the series test can reveal,
it is sufficient to restrict ourselves to the isotropic cubic Ising model
($J_1=J_2=J_3=J$, $K=\beta J=J/k_{\mathrm{B}}T$).
For that case Zhang expands his ``putative'' exact partition
function per site in (A13) on page 5406 of \cite{Z} as
\begin{equation}
Z^{1/N}=2\cosh^3K\Big[1+\frac{7}{2}\kappa^2+\frac{87}{8}\kappa^4+
\frac{3613}{48}\kappa^6+\cdots\Big],
\quad\kappa=\tanh K.
\label{foe}
\end{equation}
This differs from the well-known high-temperature series given in 
(A12) of \cite{Z} as

\begin{equation}
Z^{1/N}=2\cosh^3K\Big[1+0\kappa^2+3\kappa^4+
22\kappa^6+\cdots\Big].
\label{fo}
\end{equation}
Zhang then suggests that both results are correct, the first for
finite temperatures, the second only for an infinitesimal neighborhood
of $\beta\equiv1/k_{\mathrm{B}}T=0$,
see e.g.\ pages 5382--5384, 5394, 5400 of \cite{Z}.
This makes no sense as both series have a finite radius of
convergence. It can be rigorously shown that the first result (\ref{foe}) is wrong,
whereas the first so many terms of series (\ref{fo}), known for over 60
years \cite{DG3}, are correct in the thermodynamic limit. We shall discuss
that in a later section.

From Eq.\ (103) on page 5342 of \cite{Z} we obtain the ``putative''
low-temperature expansion of the spontaneous magnetization,
\begin{equation}
I=1-6x^{ 8}-12x^{10}-18x^{12}+\cdots,\quad x={\mathrm{e}}^{-2K}.
\label{moe}
\end{equation}
Some coefficients related to the well-known expansion, existing for
over six decades in the literature \cite{DG3}, are listed in Table 2 on
page 5380 of \cite{Z},
implying
\begin{equation}
I=1-2x^{6}-12x^{10}+14x^{12}-\cdots.
\label{mo}
\end{equation}
The textbook derivation starts with a $d$-dimensional hypercubic
lattice of $N$ sites, each site having $2d$ neighbors.
There is one state with all spins up and energy $E_+$,
$N$ states with only one spin down and energy $E_++4d J$,
etc. Thus,
\begin{equation}
I=\frac{Z_{\sigma}}{ Z}=\frac{1+(N-2)x^{2d}+\cdots}{1+Nx^{2d}+\cdots}=
1-2x^{ 2d}+\cdots,\quad 2d=6.
\end{equation}
Only if the down spin is at the position of the spin operator we get
a minus sign, or $N-2=(N-1)(+1)+(-1)$. It is well known that in the linked
cluster expansion \cite{DG3} the $N$ dependence cancels in all orders.
Again, Zhang's result (\ref{moe}) is clearly wrong, displaying $d=4$ behavior.

Much more can be said on the spontaneous magnetization and the
pair correlation. However, here it suffices to discuss errors in the treatment 
of the free energy in \cite{Z,ZM}, as the rest of Zhang's work is built on the
same erroneous foundation.

\subsection{Two different expansions for the same?}

It has been shown already in the 1960s that the high-temperature
series of $\beta f$, ($\beta f=-\lim\ln Z^{1/N}$), and all correlation functions
on the cubic lattice have finite nonzero radius of convergence.
By duality with the Ising model with 4-spin interactions on
all faces of the cubic lattice and at high temperatures, the low-temperature
series for the spontaneous magnetization $I$ should also converge.
Therefore, one cannot have two different expansions as given by Zhang.

To get out of this dilemma, Zhang (on pages 5381, 5383 and 5394
of \cite{Z}) comes with a mathematically absurd suggestion implying
that the old series (\ref{fo}) and (\ref{mo}) are asymptotic
with zero radius of convergence, whereas his ``putative solution'' is
analytic with finite radius of convergence. On pages 12 and 13 of \cite{ZM}
(and elsewhere) he claims that there is an (essential?) singularity at
$\beta=h=0$ and that this is due to the zeros of $Z^{-1}$ and that
 Perk ``went on perpetrating the fraud, discussing the singularity of
 $\beta f_N$'' \cite[page 12]{ZM} and \cite[page 63]{Zh2}, not $f$.
 To use the word ``fraud'' is clearly highly unprofessional.
 We discuss this next.

\subsection{Irrelevant pole of \em f}

It has been rigorously proved (one proof discussed in the next section)
that $\beta f$ has an absolutely convergent series, uniformly convergent
in the thermodynamic limit, so that
\begin{equation}
\beta f=\sum_{i=0}^{\infty}a_i\beta^i,\quad
|\beta|<r.
\label{bfs}
\end{equation}
Therefore,
\begin{equation}
f=\frac{a_0}{\beta}+
\sum_{i=1}^{\infty}a_i\beta^{i-1},\quad
0<|\beta|<r,
\end{equation}
is a convergent Laurent series, totally equivalent to (\ref{bfs}).

The pole at $\beta=0$ has no significance, as $\beta f$ is the relevant
quantity from statistical mechanics point of view, entering the normalization
$Z={\mathrm{e}}^{-N\beta f}$ for the Boltzmann--Gibbs canonical
distribution. Also, note that in the appendix of \cite{Z} Zhang expanded
$Z^{1/N}={\mathrm{e}}^{-\beta f}$. This makes his objection to expanding
$\beta f$ rather out of place.


\subsection{Zeros of 1/{\em Z} are irrelevant}

Zhang and March repeatedly claim the importance of the zeros
of $Z^{-1}$, with $Z=z^N$ the total partition function, see \cite[pages 64--65]{Zh2},
\cite[page 87]{Zh3} and \cite[pages 12--13]{ZM}.
However, unlike the complex Yang--Lee zeros of $Z$,
the zeros of $Z^{-1}$ are irrelevant:
\begin{itemize}
\item For a finite number $N$ of sites, $Z$ is a finite Laurent polynomial
in ${\mathrm{e}}^K$, $K\equiv\beta J$, and only can become
infinite when ${\mathrm{Re}}\,K=\pm\infty$, i.e.\ zero-temperature type limits.
\item
For the infinite system, $N\to\infty$, and finite $K$, the infinity of $Z$
should be seen as just a manifestation of the {\em thermodynamic limit}, in which
$z=Z^{1/N}$ remains finite.
\end{itemize}
One can easily see that $\beta f<0$ for $K$
real, so that $Z={\mathrm{e}}^{-N\beta f}=\infty$ for all real $K$ when
$N=\infty$. There is nothing special about the $N=\infty$ zeros of $Z^{-1}$.
Hence, Zhang and March cannot claim that the zeros of $Z^{-1}$ play
any role. They are there the same way for the one-dimensional Ising model,
$Z\approx(2\cosh K)^N$, and free Ising spins in field $B$,
$Z=(2\cosh \beta B)^N$, for $N\to\infty$.


\section{Analyticity properties of the correlation functions at high temperatures}

There are several rigorous proofs in the literature showing that the
convergence toward the thermodynamic limit is uniform and that the
resulting high-temperature series has a nonzero radius of convergence,
which then rigorously shows that the main ``putative'' results of \cite{Z}
are wrong \cite{Wu1,Perk1}. The search of such proofs was initiated by
Groeneveld \cite{Groen}, who found such a proof for the Mayer expansion
as part of his thesis research under Professor Jan de Boer, the father of
Professor Frank de Boer---one of the supervisors of Zhang Zhi-Dong's
thesis research.

Such proofs of analyticity  mostly belong to two classes, i.e.\ starting
from (linked) cluster expansions \cite{DG3}, like the Mayer expansion, or
applying linear correlation-function identities \cite{Ruelle}.
In the next few subsections we shall present an outline of the proof
in \cite{P1}, which belongs to the second class. The intermediate steps
in this proof can be used to pinpoint other errors in the works of
Zhang and March.

We first prove that the correlation functions of the Ising model on the
periodic cubic lattice of size $N=n^3$, have high-temperature series
absolutely convergent when $|\beta J|<r_0$, uniform in $n$. 
The corresponding analyticity statement for $\beta f_N$ then follows using
the elementary identity $\frac{\partial(\beta f_N)}{\partial\beta}=u_N=
3J\langle\sigma\sigma'\rangle_N$, with $\sigma$ and $\sigma'$
nearest-neighbor spins. As more and more coefficients become
independent of $n$, the analyticity then carries over in the
thermodynamic limit $N\to\infty$.

\subsection{Laurent Polynomial Lemma\label{Laur}}

The partition function $Z_N$ of the Ising model on a periodic cubic
lattice of sides $n$ lattice spacings, $N=n^3$, is a Laurent polynomial
in ${\mathrm{e}}^{\beta J}$. This means that it is a polynomial
in both ${\mathrm{e}}^{\beta J}$ and ${\mathrm{e}}^{-\beta J}$.
Therefore, $\beta f_N=-N^{-1}\ln Z_N$ is singular only
for the zeros of this Laurent polynomial and for some cases with
$|{\mathrm{e}}^{\pm\beta J}|=\infty$. As $Z_N$ is a sum of positive
terms for real $\beta J$, 
{\em $Z_N$ cannot have zeros on the real $\beta J$-axis}.

It follows that all correlation functions $\langle\prod_l\sigma_l\rangle_N$
are {\bf meromorphic} functions with poles only at the zeros of $Z_N$.
Hence,
\begin{equation}
\Big\langle\prod_l\sigma_l\Big\rangle_N=\sum_{i=1}^{\infty}c_i(\beta J)^i,
\quad |c_i|<C_N\,r^{-i},
\end{equation}
with $r$ the absolute value of the zero closest to $\beta J=0$
and $C_N$ some positive constants.

\subsection{Lemma (M. Suzuki, 1965) for {\em B} = 0}

\begin{equation}
\bigg\langle\prod_{i=1}^m\sigma_{j_i}\bigg\rangle_N=
\frac1m\sum_{k=1}^m
\bigg\langle\bigg(\prod_{i=1, i\ne k}^m\sigma_{j_i}\bigg)
\tanh\bigg(\beta J\!\sum_{l\;{\mathrm{nn}}\;j_k}\!
\sigma_{l}\bigg)\bigg\rangle_N,
\end{equation}
where $j_1,\ldots,j_m$ are the labels of $m$ spins and $l$
runs through the labels of the six spins that are nearest
neighbors of $\sigma_{j_k}$ \cite{P1,Suz,Suz2}. The averaging
over $k$ treats the $m$ spins  $\sigma_{j_k}$ symmetrically.
The lemma without the averaging over $k$ was presented
by Suzuki and can be used instead, selecting one of the $m$
spins at random.

\subsection{Expanding tanh Lemma}

It is straightforward \cite{P1} to prove the following or to verify
it using Mathematica or Maple:
\begin{equation}
\tanh\bigg(\beta J\sum_{l=1}^6\sigma_l\bigg)=
a_1\sum_{(6)}\sigma_l
+a_3\sum_{(20)}\sigma_{l_1}\sigma_{l_2}\sigma_{l_3}+a_5
\sum_{(6)}\sigma_{l_1}\sigma_{l_2}\sigma_{l_3}\sigma_{l_4}\sigma_{l_5},
\end{equation}
where the sums are over the 6, 20, or 6 choices of choosing 1, 3, or 5
spins from the given $\sigma_1,\ldots,\sigma_6$. The coefficients $a_i$ are
\begin{eqnarray}
&&\displaystyle
a_{1,5}=\frac1{24}\Big(\frac{p_1t}
{1+(p_1t)^2}+\frac{p_2t}{1+(p_2t)^2}\Big)
\pm\frac{\sqrt{2}}{8}\Big(\frac{p_3t}
{1+(p_3t)^2}+\frac{p_4t}{1+(p_4t)^2}\Big)
+\frac13\,\frac{p_5t}{1+(p_5t)^2},\nonumber\\
&&\displaystyle
a_{3\phantom{,5}}=
\frac1{24}\Big(\frac{p_1t}{1+(p_1t)^2}+\frac{p_2t}{1+(p_2t)^2}\Big)
-\frac16\,\frac{p_5t}{1+(p_5t)^2},\nonumber\\
&&\displaystyle
p_{1,2}=2\pm\sqrt{3},\quad p_{3,4}=\sqrt{2}\pm1,\quad p_5=1,
\qquad (p_1p_2=p_3p_4=1).
\end{eqnarray}
The poles of the $a_i$ are at $t=\pm(2\pm\sqrt{3}){\mathrm{i}}$,
$t=\pm(\sqrt{2}\pm1){\mathrm{i}}$,
and $t=\pm{\mathrm{i}}$. The $a_i$ have series
expansions in terms of odd powers of $t$ alternating
in sign and converging absolutely
as long as $|\beta J|<\arctan(2-\sqrt{3})=\pi/12$.


\subsection{Uniform convergence for finite \em N}

Expanding the tanh in Suzuki's lemma replaces any even correlation by a linear
combination of even correlations, with coefficients given by the $a_i$. As
for fixed $|t|\equiv x$, each $|a_i|$ is maximal for imaginary $t={\mathrm{i}}x$,
we find for the sum $s$ of the absolute values of all coefficients:
\begin{equation}
s\equiv6|a_1|+20|a_3|+6|a_5|\le
\frac{2x(3-x^2)(1-3x^2)}{(1-x^2)(1-14x^2+x^4)}.
\end{equation}
It easily follows that $s<1$ for
\begin{eqnarray}
&&|t|<(\sqrt{3}-\sqrt{2})(\sqrt{2}-1)=t_0=0.131652497\cdots,
\quad\hbox{or}\nonumber\\
&&|\beta J|<\arctan[(\sqrt{3}-\sqrt{2})(\sqrt{2}-1)]
=0.130899693\cdots.
\label{bound}
\end{eqnarray}

This determines a lower bound on the radius of convergence: 
Keep repeatedly applying Suzuki's lemma on all correlations
that are not the ``zero-point correlation function''
$\langle1\rangle=1$. Terms with $\langle1\rangle$ are explicitly
known and terms that are not are at least one order higher in $t$
or $\beta J$. After one iteration we have the very generous
upper bound
\begin{equation}
\bigg|\bigg\langle\prod_{i=1}^m\sigma_{j_i}\bigg\rangle_{\!N}\bigg|<s+
s\max\bigg|\bigg\langle\prod_{i=1}^{m'}\sigma_{j'_i}\bigg\rangle_{\!N}\bigg|,
\label{ineq}
\end{equation}
with maximum over all correlations generated by Suzuki's lemma.
After $I$ iterations, a given even correlation
is then split into an explicitly given sum bounded by $s+s^2+\cdots+s^I$
and a remainder sum with correlations left for further iteration,
higher order in $t$ or $\beta J$. If we increase the right-hand side
of (\ref{ineq}) by taking the maximum $M$ over all $2^N-1$
correlation functions that are not $\langle1\rangle$ ($m\ne0$), then it
immediately follows that $M<s+sM$, so that
\begin{equation}
\bigg|\bigg\langle\prod_{i=1}^m\sigma_{j_i}\bigg\rangle_{\!N}\bigg|<
\frac{s}{1-s},\quad\mbox{if}\quad m\ne0,\quad |t|<t_0.
\end{equation}
This implies that the remainder term is bounded by $s^{I+1}/(1-s)$.
This is consistent with iterating ad infinitum, which gives
$s^{I+1}+s^{I+2}+\cdots=s^{I+1}/(1-s)$. It is also consistent with the
fact that each correlation function can have at most poles given
by the complex Yang--Lee zeros of $Z_N$.

As the bounds are independent of $N$, we have shown absolute
and uniform convergence of the high-temperature series for any
correlation function with finite radius of convergence bounded
below by (\ref{bound}).

\subsection{Uniform convergence for infinite \em N}

As a consequence, $\langle\prod_{i=1}^m\sigma_{j_i}\rangle_N$
converges to a unique limit as $N\to\infty$ for $|t|<2-\sqrt{3}$, defined
by its series. Let $d$ be the largest edge of the minimal parallelepiped
containing all sites $j_1,\ldots,j_m$. Then the coefficient of $t^k$
with $k<n-d$,  for the cubic lattice with $N=n^3$ sites and periodic
boundary conditions, equals the corresponding coefficient for
larger $N$. It takes at least $n-d$ iteration steps to notice the finite
size of the lattice. Increasing $N$ by one step makes one more
coefficient independent of $N$. As $N\to\infty$ all coefficients take
their thermodynamic limit value and the remainder term tends to
zero uniformly, according to the previous subsection.

\subsection{Theorem for reduced free energy and its thermodynamic limit}

The reduced free energy $\beta f_N$ for arbitrary $N$ and its
thermodynamic limit $\beta f$ are analytic in $\beta J$ for sufficiently
high temperatures. They have series expansions in $t$ or $\beta J$
with radius of convergence bounded below by (\ref{bound}) and uniformly
convergent for all $N$ including $N=\infty$. The first $n-1$ coefficients
of these series for $N=n^3$ equal their limiting values for $N=\infty$.

This is easily proved using
\begin{equation}
u_N=\frac{1}{N}\langle{\cal H}_N\rangle_N^{\phantom{y}}=
\frac{\partial(\beta f_N)}{\partial\beta}=
-3J\langle\sigma_{0,0,0}\sigma_{1,0,0}\rangle_N^{\phantom{y}},
\end{equation}
with $\sigma_{0,0,0}$ and $\sigma_{1,0,0}$ a nearest-neighbor pair of
spins. The proof then follows from a well-known theorem in
complex calculus integrating the series for $u_N$ and
using $Z_N|_{\beta=0}=2^N$,
implying $\lim_{\beta\to0}\beta f_N=-\log2$.

\section{The first nontrivial coefficient}

Applying Suzuki's lemma once to $\langle\sigma_{0,0,0}\sigma_{1,0,0}\rangle$,
we find
\begin{equation}
\langle\sigma_{0,0,0}\sigma_{1,0,0}\rangle=a_1\,\langle1\rangle+{\mathrm{O}}(t^2)
=t+{\mathrm{O}}(t^2)=\beta J+{\mathrm{O}}(\beta^2).
\end{equation}
Hence,
\begin{equation}
\beta f=-\log2-\int_0^{\beta}3J\langle\sigma_{0,0,0}\sigma_{1,0,0}\rangle
{\mathrm{d}}\beta=-\log2-\frac32(\beta J)^2+{\mathrm{O}}(\beta^3).
\end{equation}
This and the next few terms so obtained agree with the usual series
expansion (\ref{fo}), but disagree with Zhang's ``putative" exact result.

The only possible conclusion is that the conjectured answers of Zhang \cite{Z,ZM}
are wrong and we shall next see more reasons why.

\section{No obvious choice of weight functions}

One problem with conjecture 2 \cite{Z,ZM} is that there is no obvious choice
for the weight functions. In (49) on page 5325 of \cite{Z} Zhang writes
\begin{eqnarray}
\displaystyle
N^{-1}\ln Z&=&\displaystyle\ln2+\frac1{2(2\pi)^4}
\int_{-\pi}^{\pi}\int_{-\pi}^{\pi}\int_{-\pi}^{\pi}\int_{-\pi}^{\pi}
\ln\big[\cosh 2K\cosh2(K'+K''+K''')\nonumber\\
&&-\sinh2K\cos\omega'-\sinh2(K'+K''+K''')(w_x\cos\omega_x\nonumber\\
&&+w_y\cos\omega_y+w_z\cos\omega_z)\big]d\omega'd\omega_xd\omega_yd\omega_z,
\end{eqnarray}
with $K=K'=K''=K'''$ for the isotropic cubic lattice.

Even though this is a result of flawed assumptions, it is an integral
transform that could give the right answer with a suitable choice of
weight functions $w_x$, $w_y$, $w_z$. 
Zhang's wrong ``putative'' result comes from the choice
$w_x=1$, $w_y=w_z=0$. In appendix A of \cite{Z} Zhang reversely engineers
an other (truncated) series choice derived from known coefficients of
the Domb--Sykes high-temperature series. There is no more information
than is in the limited series results provided by others, so that this
is not an exact result.

\section{Conjecture 1 is manifestly wrong}

The original paper \cite{Z} has an error in the application
of the Jordan--Wigner transformation pointed out in \cite{Perk1}.
This error has only been corrected explicitly in \cite{ZM},
which makes it easy to pinpoint the error with Conjecture 1:
Zhang and March violate the property that Lie groups are closed
under commutation and that the product of fermionic
Gaussians is another fermionic Gaussian \cite{P3}.

It is well-known that in the spinor representation of the orthogonal
groups each element $g$ can be written as a ``fermionic Gaussian'' of
the form
\begin{equation}
g=\exp\bigg(\frac12\sum_i\sum_j A_{ij}\bG_i\bG_j\bigg),
\quad A_{ji}=-A_{ij},
\end{equation}
with Clifford algebra elements satisfying
$\bG_i\bG_j+\bG_j\bG_i=2\delta_{ij}$ and antisymmetric complex
coefficients $A_{ij}$. The spinor representation has been used in the
Ising context first by Kaufman \cite{Kaufman} in 1949.

The closure property of Lie groups and the Baker--Campbell--Hausdorff
formula require that any product, commutator or inverse of elements of this form
is again of the same fermionic Gaussian form. Equivalently, the sum, commutator,
or inverse of Lie algebra elements can only produce Lie algebra elements.

\subsection{Remark: Lie group and Lie algebra}

The group elements $g$ act on the $\bG\,$'s as
\begin{equation}
\bG_k\quad \longrightarrow\quad g\bG_k g^{-1}.
\end{equation}
Choosing infinitesimal
\begin{equation}
g={\bf1}+\frac{\varepsilon}2\sum_i\sum_j A_{ij}\bG_i\bG_j
+{\mathrm{O}}(\varepsilon^2),
\label{g}\end{equation}
we find from (\ref{g}) in ${\mathrm{O}}(\varepsilon^2)$, that the corresponding Lie algebra action is
\begin{equation}
\bigg[\frac12\sum_i\sum_j A_{ij}\bG_i\bG_j\,,\,\bG_k\bigg]=
\sum_i A_{ik}\bG_i\,,
\end{equation}
showing that the infinitesimal action is through multiplication
by antisymmetric matrices, the generators of rotations.

Therefore, the $g$'s indeed form a representation of a rotation group.

\subsection{Transfer matrix}

It is easily proved that the free energy per site of the
ferromagnetic Ising model in the thermodynamic limit
does not depend on boundary conditions. Therefore,
the Hamiltonian (1) in \cite{Z} can be rewritten using the
scew boundary conditions of Kramers and Wannier \cite{KW} as
\begin{equation}
-\beta\hat H=\sum_{\tau=1}^n\sum_{j=1}^{ml}
\Big[K s_j^{(\tau)}s_j^{(\tau+1)}+
K's_j^{(\tau)}s_{j+1}^{(\tau)}+
K''s_j^{(\tau)}s_{j+m}^{(\tau)}\Big].
\end{equation}
For this purpose we have made the change of notation
\begin{equation}
s_{\rho,\delta}^{(\tau)}\equiv s_j^{(\tau)},
\qquad j\equiv\rho+(m-1)\delta,
\qquad\sum_{\rho=1}^m\sum_{\delta=1}^l=\sum_{j=1}^{ml},
\end{equation}
where $\tau=1,\cdots,n;\quad\rho=1,\cdots,m;\quad\delta=1,\cdots,l$.

This leads to the transfer matrix $\Tb=\Vb_3\Vb_2\Vb_1$ in \cite{ZM}, with
\begin{eqnarray}
&&\displaystyle
\Vb_3=\exp{\bigg(-{\mathrm{i}}K''\sum_{j=1}^{ml}
\bG_{2j}\bigg[\prod_{k=j+1}^{j+m-1}{\mathrm{i}}\bG_{2k-1}\bG_{2k}
\bigg]\bG_{2j+2m-1}\bigg)},\label{V3}\\
&&\displaystyle
\Vb_2=\exp{\bigg(-{\mathrm{i}}K'\sum_{j=1}^{ml}
\bG_{2j}\bG_{2j+1}\bigg)},\quad
\Vb_1=\exp{\bigg({\mathrm{i}}K^{\ast}\sum_{j=1}^{ml}
\bG_{2j-1}\bG_{2j}\bigg)},
\label{V12}\end{eqnarray}
compare (15), (16) and (17) of \cite{ZM}, with $n$ replaced by $m$.
Clearly, (\ref{V3}) is not of the fermionic Gaussian form and,
therefore, not an element of the group.

 At this point, Zhang introduces a fourth dimension, stacking $o$
copies of the model. Without changing the free energy per site
in the large system limit, one can connect the copies to give (\ref{V3})
and (\ref{V12}) with the upper bounds of the sums $ml$ replaced by $mlo$.

Next, Zhang made the absurd conjecture that multipying
$\Vb_3{}'\Vb_2{}'\Vb_1{}'$ so obtained by
\begin{equation}
\Vb_4{}'=\exp\bigg\{-{\mathrm{i}}K'''\sum_{j=1}^{mlo}\bG_{2j}\bG_{2j+1}\bigg\},
\qquad K'''=\frac{K'K''}{K},
\end{equation}
as given in (18) and (19) in \cite{ZM}, miraculously produces a rotation
group element. This is in violation of the Baker--Campbell--Hausdorff formula.
The argumentation in \cite{Z} for the form of $K'''$ is ad hoc and also
makes no sense.

\section{Some other issues out of many more}

Zhang and March also falsely claim that setting $\beta=1$ in 1960's
references is a loss of generality, losing $T=\infty$. On the contrary \cite{P2},
as $\beta f$ is only a function of $K=\beta J$, this is no problem.
Having $J\equiv K$ and choosing a fixed new $\bar J$ and a new
$\bar\beta=J/\bar J\not\equiv1$, we can write $J=K=\bar\beta\bar J$,
recovering the full  two-variable case depending on $\beta$
(including $\beta=0$) and a new $J$ (omitting the bars) \cite{P3}.

Next, Zhang and March fail to realize that $K_{\beta\phi'}(X,T)$
in \cite{GMR} vanishes for $\beta=0$. The
cited inequality does not fail for $\beta=0$ \cite{P3}.

Also, the three-dimensional Virasoro algebra approach in \cite{ZM,Zh1}
is based on an erroneous solution of the 3D Ising model and twice
writing ${\mathrm{Re}}|{\mathrm{e}}^{{\mathrm{i}}\phi_i}|$, the real
part of a positive real number \cite[pages 39--40]{Zh1}, is another
error \cite{P3}.

\section{Conclusion}

In conclusion, much of what Zhang and March wrote about the
three-dimensional Ising model is either misleading or completely wrong.
All their main results are in error. It is said in \cite{ZM} that everything is
based on two conjectures. This is also false, as a careful reading of \cite{Z}
reveals that many steps there lack mathematical logic and should be
considered unfounded assumptions. However, there should be no need
to go through these other issues in lengthy detail, after all that is already said.



\begin{thebibliography}{99}
\itemsep=-4pt plus.2pt minus.2pt  
\small
\bibitem{Z}
Zhang Z-D 2007
\textit{Philos. Mag.} {\bf 87} 5309--5419,
arXiv:0705.1045 (176 pp)
\bibitem{Wu1}
Wu F Y, McCoy B M, Fisher M E and Chayes L 2008
\textit{Philos. Mag.} {\bf 88} 3093--3095, arXiv:0811.3876
\bibitem{Perk1}
Perk J H H 2009
\textit{Philos. Mag.} {\bf 89} 761--764, arXiv:0811.1802
\bibitem{Z2}
Zhang Z-D 2008
\textit{Philos. Mag.} {\bf 88} 3097--3101, arXiv:0811.2330
\bibitem{ZM} 
Zhang Z-D 2013
\textit{Chin. Phys.} B {\bf 22} 030513 (15 pp),
arXiv:1305.2956 (53 pp)
\bibitem{DG3}
Domb C and Green M S editors 1974
\textit{Series Expansions for Lattice Models},
{Phase Transitions and Critical Phenomena} (Vol. 3)
(London: Academic Press)
\bibitem{Zh2}
Zhang Z-D and March N H 2012
\textit{Bull. Soc. Sci. Lettres  \L\'od\'z} S\'er. Rech. D\'eform.
{\bf 62}:3 61--69, arXiv:1209.3247
\bibitem{Zh3}
Zhang Z-D and March N H 2013
\textit{Bull. Soc. Sci. Lettres  \L\'od\'z} S\'er. Rech. D\'eform.
{\bf 63}:1 85--88, arXiv:1209.3247
\bibitem{Groen}
Groeneveld J 1962
\textit{Phys. Lett.} {\bf 3} 50--51
\bibitem{Ruelle}
Ruelle D 1969
\textit{Statistical Mechanics, Rigorous Results}
(New York: Benjamin)
\bibitem{P1}
Perk J H H 2012
\textit{Bull. Soc. Sci. Lettres  \L\'od\'z} S\'er. Rech. D\'eform.
{\bf 62}:3 45--59, arXiv:1209.0731
\bibitem{Suz} 
Suzuki M 1965
\textit{Phys. Lett.} {\bf 19} 267--268
\bibitem{Suz2} 
Suzuki M 2002
\textit{Intern. J. Mod. Phys.} B {\bf 16} 1749--1765
\bibitem{P3}
Perk J H H 2013
\textit{Bull. Soc. Sci. Lettres  \L\'od\'z} S\'er. Rech. D\'eform.
{\bf 63}:1 89--92, arXiv:1209.0731
\bibitem{Kaufman} 
Kaufman B 1949
\textit{Phys. Rev.} 76 1232--1243.
\bibitem{KW} 
Kramers H A and Wannier G H 1941
\textit{Phys. Rev.} {\bf 60} 252--262
\bibitem{P2}
Perk J H H 2012
\textit{Bull. Soc. Sci. Lettres  \L\'od\'z} S\'er. Rech. D\'eform.
{\bf 62}:3 71--74, arXiv:1209.0731
\bibitem{GMR}
Gallavotti G, Miracle-Sol\'e S and Robinson D W 1967
\textit{Phys. Lett.} A {\bf 25} 493--494
\bibitem{Zh1}
Zhang Z-D and March N H 2012
\textit{Bull. Soc. Sci. Lettres  \L\'od\'z} S\'er. Rech. D\'eform.
{\bf 62}:3 35--44, arXiv:1110.5527
\end{thebibliography}
\end{document}